\documentclass[aps,prl,twocolumn,showpacs,10pt,amsmath,footinbib]{revtex4-1}
\usepackage{graphicx,color}
\usepackage{dcolumn}
\usepackage{bm}
\usepackage{hyperref}
\usepackage{psfrag}

\newcommand{\bV}{{\bf V}}
\newcommand{\bF}{{\bf F}}
\newcommand{\oV}{{\mbox{$\overline{V}$}}}
\newcommand{\ov}{{\mbox{$\overline{v}$}}}
\newcommand{\oL}{{\mbox{$\overline{L}$}}}

\begin{document}
\preprint{APS/123-QED}
\title{Regimes of turbulence without an energy cascade}
\author{C. F. Barenghi$^1$}
\email{carlo.barenghi@newcastle.ac.uk}
\author{Y. A. Sergeev$^2$}
\author{A. W. Baggaley$^1$}
\affiliation{$^1$ Joint Quantum Centre (JQC) Durham-Newcastle, 
School of Mathematics and Statistics, 
Newcastle University, Newcastle upon Tyne, NE1 7RU, UK \\
$^2$ Joint Quantum Centre (JQC) Durham-Newcastle, School of Mechanical and Systems Engineering, 
Newcastle University, Newcastle upon Tyne, NE1 7RU, UK}

\date{\today}%

\begin{abstract}
Experiments and numerical simulations of turbulent $^4$He and $^3$He-B
have established that, at hydrodynamic length scales larger than the
average distance between quantum vortices, the
energy spectrum obeys the same 5/3 Kolmogorov law which is observed
in the homogeneous isotropic turbulence of ordinary fluids. The
importance of the 5/3 law is that it points to the existence of a
Richardson energy cascade from
large eddies to small eddies.
However, there is also evidence of quantum turbulent regimes without
Kolmogorov scaling. This raises the important questions
of why, in such regimes,
the Kolmogorov spectrum fails to form, what is the physical nature
of turbulence without energy cascade, and whether hydrodynamical models
can account for the unusual behaviour of turbulent superfluid helium.
In this work we describe simple physical mechanisms which prevent the
formation of Kolmogorov scaling in the thermal counterflow, 
and analyze the conditions necessary for emergence of quasiclassical 
regime in quantum turbulence generated by injection of vortex rings at 
low temperatures. Our models justify the hydrodynamical description 
of quantum turbulence and shed light into an unexpected regime 
of vortex dynamics.
\end{abstract}

\maketitle

\section*{Introduction}

Unlike what happens in ordinary fluids, vorticity in quantum fluids 
is constrained to vortex filaments of quantized circulation.
Recent experimental, numerical and theoretical 
studies \cite{Barenghi-2014-spectra,Nore-1997,Baggaley-Laurie} have revealed 
the surprising result that quantum turbulence, 
despite the discrete nature of the vorticity, 
obeys the same Kolmogorov scaling which is observed for homogeneous 
isotropic turbulence in ordinary fluids. More precisely, the superfluid 
kinetic energy spectrum scales as $E_k \sim k^{-5/3}$  in the 
`hydrodynamical' range $k_D = 2 \pi/D \ll k \ll k_{\ell} = 2 \pi/\ell$, 
where $k$ is the wavenumber, $D$ the size of the system, and $\ell$ the 
average distance between vortex lines. 
(What happens at length scales smaller than  $\ell$ is very 
interesting  - issues which are debated are the Kelvin wave cascade and the 
possible existence of an energy flux bottleneck -  but is outside the 
scope of this work.) Evidence for Kolmogorov scaling has been found in 
both bosonic $^4$He and in fermionic $^3$He-B, at both high temperatures 
(where helium acquires a two-fluid nature due to the presence of the normal 
fluid) and low temperatures. It has also been found, 
numerically~\cite{Baggaley-2011-stats} and 
experimentally~\cite{LaMantia-2014-stats}, that, when averaged over scales 
larger than $\ell$, the turbulent velocity components obey the same 
Gaussian statistics of classical turbulence. The 
current~\cite{Barenghi-2014-intro} interpretation of these results is that 
quantum turbulence represents the `skeleton' of ordinary turbulence.

However, there is also evidence for a very different spectral nature, 
in which the largest eddies are weak, most of the energy is contained in 
the intermediate scales, and the large wavenumber range has the $k^{-1}$ 
dependence of isolated vortex lines. Such features suggest a tangle of 
randomly oriented vortex lines whose velocity fields tend to cancel 
each other out. This second form of turbulence, named `Vinen' 
or `ultraquantum' turbulence to distinguish it from the previous 
`Kolmogorov' or `quasiclassical' turbulence \cite{Volovik}, has been 
identified both numerically \cite{Baggaley-2012-ultra} and 
experimentally (in low temperature $^4$He \cite{Walmsley-2008,Zmeev2015} and in 
$^3$He-B \cite{Bradley2006}), and also (numerically) in high temperatures 
$^4$He driven by a heat current \cite{Sherwin2012} (thermal counterflow).

The natural question is whether either Vinen turbulence is some new form 
of disorder (if so, why it has not been observed in classical turbulence?) 
or there are physical mechanisms which prevent the development of the 
classical Kolmogorov spectrum and the energy cascade. In this report 
we shall argue that the latter is the solution of this important puzzle.

\section*{Kolmogorov scenario}

In this section we briefly summarize the Kolmogorov phenomenology 
for classical turbulence (undoubtedly, the content of this section 
is well known and can be found in every standard text on 
classical turbulence~\cite{Frisch}). 
This will make our derivations of the conditions for/against the 
existence of the kinetic energy cascade and of the associated 
Kolmogorov scaling in quantum turbulence more transparent.

In classical homogeneous and isotropic turbulence (away from the 
boundaries) the Richardson energy cascade takes place if there exists 
an interval of length scales (known as the `inertial range') such that, 
at every scale $r$ within this range, the dissipation 
time $\tau_{\rm d} \approx r^2/\nu$ (where $\nu$ is the kinematic 
viscosity) is much longer than the eddie turnover 
time $\tau_{\rm r} \approx r/u_r$:
\begin{equation}
\tau_{\rm d}\gg\tau_r\,,
\label{eq:tau}
\end{equation}
with the velocity at the length scale $r$ given by
\begin{equation}
u_r=\epsilon^{1/3} r^{1/3},
\label{eq:ur}
\end{equation}
where $\epsilon=-dE/dt$ is the energy dissipation rate and $E$ 
the energy per unit mass (we assume that the fluid has constant density).

Condition~(\ref{eq:tau}) for the existence of the cascade can be 
reformulated as ${\rm Re}_r\gg1$, where
\begin{equation}
{\rm Re}_r=\frac{u_r r}{\nu}=\frac{\epsilon^{1/3}r^{4/3}}{\nu}
\label{eq:Re}
\end{equation}
is the scale-by-scale Reynolds number. Inequality~(\ref{eq:tau}) 
becomes invalid at the Kolmogorov length scale, $r=\eta=\nu/u_r$, 
where, by definition, ${\rm Re}_r=1$ (that is, viscous and inertial 
forces are comparable).

The energy spectrum, $E_k$, is defined by
\begin{equation}
E=\frac{1}{V}\int\frac{u^2}{2}\,dV=\int E_k\,dk\,,
\label{eq:Ek}
\end{equation}
where $V$ is the volume. Within the inertial range, the Kolmogorov 
scaling of the energy spectrum is obtained assuming that $E_k$ 
depends only on $\epsilon$ and $k$. Simple dimensional analysis then 
yields the famous result that, within the inertial range 
$k_D \ll k \ll k_{\eta}= 2 \pi/\eta$, the energy spectrum is
\begin{equation}
E_k=C_K\epsilon^{2/3}k^{-5/3}\,,
\label{eq:5/3}
\end{equation}
where $C_K$ is a dimensionless constant of order unity.

\section*{Results}

\subsection*{Hydrodynamic regime of quantum turbulence}

We now turn to quantum turbulence. Unlike classical turbulence, where 
vorticity is continuous and eddies have arbitrary shape and strength, 
quantum turbulence consists of individual vortex lines. Each vortex 
line carries one quantum of circulation $\kappa=h/m$, where $h$ is 
Planck's constant and $m$ the mass of the relevant boson (one atom 
for $^4$He, one Cooper pair for $^3$He-B).

In analogy with classical turbulence, we are interested in the 
existence of a Richardson energy cascade in the hydrodynamic range 
$k_D \ll k \ll k_{\ell}$, which is the regime of vortices interacting 
with each other; we are not interested in the $k \ge k_{\ell}$ regime 
of isolated vortex lines (an important but different physical problem 
with no direct relation to classical fluid dynamics).

We, therefore, consider the length scales $r$ such that
\begin{equation}
r\gg\ell\,,
\label{eq:rggl}
\end{equation}
where the mean intervortex distance, $\ell$ can be inferred from the 
observed vortex line density $L$, defined as the total length of vortex 
lines per unit volume, as $\ell=L^{-1/2}$ However, in the zero-temperature 
limit this relation holds only for the smoothed line density; owing to the 
presence of high frequency Kelvin waves undamped by the mutual friction, 
in the zero-temperature limit the actual vortex line density exceeds the 
smoothed line density, 
see Ref.~\cite{VinenNiemela} for details which will also be discussed 
below in the penultimate section of this report.

Similar to the Kolmogorov phenomenology for classical turbulence, the 
conditions for existence of the Richardson cascade and, therefore, 
the $k^{-5/3}$ Kolmogorov scaling of the superfluid kinetic energy 
spectrum, can be obtained by comparing the time scale of dissipation 
with the turnover time of the macroscopic eddies. In the next two sections 
we shall make this comparison (distinguishing between high temperature 
and low temperature regimes) for turbulent states which, anomalously, 
do not follow the Kolmogorov scaling.

For the sake of brevity, we shall call {\it non-cascading} the turbulence 
which does not exhibit scale-by-scale energy transfer. Thus, for a 
homogeneous isotropic system, the energy spectrum does not scale as 
$k^{-5/3}$ at large wavenumbers $k$; a noticeable feature of such 
non-cascading turbulence is that kinetic energy is concentrated 
at some intermediate wavenumbers, 
giving the spectrum $E_k$ the shape of a `bump' followed by $k^{-1}$ 
behaviour at large $k$.

\subsection*{Non-cascading turbulence at high temperatures}

We start with quantum turbulence at high temperatures such that 
$1\,{\rm K}<T<T_\lambda$ for $^4$He, where $T_\lambda\approx2.17\,{\rm K}$ 
is the superfluid transition temperature, or $0.5T_c<T<T_c$ for $^3$He-B, 
where $T_c\approx0.9\,{\rm mK}$ is the critical temperature of the 
superfluid transition. At high temperatures the dissipation is caused by 
the mutual friction between the normal fluid and superfluid vortices. 
An anomalous, non-cascading superfluid energy spectrum has been 
predicted \cite{Sherwin2012} for $^4$He counterflow and for grid 
turbulence in $^3$He-B in the presence of very viscous, stationary normal 
fluid~\cite{Vinen2005}. Similar conclusions were reached in Refs.~\cite{Lvov2004,Lvov2006}, where it has been shown that, in the case where the normal fluid is either stationary or non-turbulent, the strong mutual friction, which is dissipative at all length scales, prevents the emergence of the inertial range and the $5/3$ energy spectrum in the superfluid component. Our aim here is to reveal the physical mechanism 
which prevents the formation of the classical Kolmogorov spectrum.

To find the timescale associated with the mutual friction, we develop the 
following simple macroscopic model of thermal counterflow in $^4$He.

The coarse-grained (that is, averaged over a scale much larger than the 
intervortex distance) superfluid and normal fluid velocities, $\bV_s$ 
and $\bV_n$, are governed by the equations~\cite{BDV1983,RocheHVBK}
\begin{eqnarray}
\rho_s \frac{D \bV_s}{Dt} 
&=& -\frac{\rho_s}{\rho} \nabla p + \rho_s S \nabla T -\bF\,,
\label{eq:Vs} \\
\rho_n \frac{D \bV_n}{Dt} 
&=& -\frac{\rho_n}{\rho} \nabla p - \rho_s S \nabla T +\bF + \mu \nabla^2 \bV_n\,,
\label{eq:Vn} \\
\nonumber
\end{eqnarray}
where $p$ is the pressure, $S$ the entropy, $T$ the temperature, $\rho_s$ and $\rho_n$ the 
superfluid and normal fluid densities, $\rho=\rho_s+\rho_n$ the total 
density, $\mu$ the viscosity, and $D/Dt$ the convective derivative. 
The mutual friction force per unit volume is
\begin{equation}
\bF=\alpha \kappa L \rho_s (\bV_s-\bV_n)\,,
\label{eq:Fr}
\end{equation}
where $\alpha$ is the temperature-dependent mutual friction coefficient 
(for $^4$He its values, obtained from the counterflow measurements, are tabulated in Ref.~\cite{RJD-CFB-data}). 
The quantum of circulation is $\kappa=0.997\times10^{-3}$.

We consider for simplicity steady one-dimensional
flow along the $x$-direction of a long channel which is closed at one 
end and open to the helium bath at the other end. At the closed end, an 
electrical resistor dissipates a given heat flux $\dot{Q}$, which is 
carried away by the normal fluid at the velocity $V_n=\dot{Q}/(\rho ST)$. 
Superfluid flows in the opposite direction to maintain the condition of 
zero mass flux, $\rho_n V_n + \rho_s V_s=0$. In this way a relative motion 
(counterflow) $V_n-V_s=V_{ns}$ is set up beween the normal fluid and the 
superfluid, which is proportional to the applied heat flux, 
$V_{ns}={\dot Q}/(\rho_s ST)$. 
The importance of this flow configuration cannot be understated, as it
is used to study the exceptional heat conducting properties of liquid
helium as a coolant in engineering applications.
Provided that $\dot Q$ exceeds a small critical value, the superfluid 
becomes turbulent, and a tangle of vortex lines fills the channel with 
vortex line density $L$.  We assume that $\dot Q$ is not so large that 
the normal fluid becomes turbulent. Let $V_s=\oV_s$, $V_n=\oV_n$ and $\oL$ 
be the values of superfluid velocity, normal fluid velocity and vortex 
line density, respectively, in the steady-state regime at given temperature 
$T$ and heat flux ${\dot Q}$.

To simplify the problem and to highlight the role of the friction, we 
neglect the viscous term in eq.~(\ref{eq:Vn}) (which would cause a 
pressure drop along the channel), and obtain the pressure and temperature 
gradients induced by the quantum turbulence:
\begin{equation}
\frac{dp}{dx}=0\,,
\label{eq:dp}
\end{equation}
and
\begin{equation}
\frac{dT}{dx}=\frac{\alpha\kappa\oL}{S}\,\oV_{ns}\,.
\label{eq:dT}
\end{equation}
Eq.~(\ref{eq:dp}) is known as the Allen-Reekie rule and has been verified
in the experiments. Experiments and 
numerical simulations show that 
\begin{equation}
\oL=\gamma^2\oV_{ns}^2\,,
\label{eq:gamma}
\end{equation}
where $\gamma$ is a temperature dependent parameter \cite{Adachi2010}; 
from eq.~(\ref{eq:dT}), this means that the temperature gradient $dT/dx$ 
is proportional to $\oV_{ns}^3$, in agreement with experiments.

The next step is consider the effect of superfluid velocity 
fluctuations $v_s$ on top of the steady flow $\oV_s$. Proceeding with a 
one-dimensional model, we write 
\begin{equation}
V_s=\oV_s +v_s(x,t)\,.
\label{eq:fluctuations}
\end{equation}

For the sake of simplicity, and, again, to bring in evidence the role 
of the mutual friction, we assume that the velocity of the normal fluid 
remains constant, and linearize the friction, neglecting fluctuations 
of $L$, $T$ and $p$, so that $\oL=L$. After subtracting the steady state, 
the superfluid equation reduces to
\begin{equation}
\rho_s \left( \frac{\partial v_s}{\partial t}
+\oV_s \frac{\partial v_s}{\partial x} \right)
=-\alpha \kappa L \rho_s v_s\,.
\end{equation}
Having assumed that the channel is long, we change to the moving reference 
frame $x'=x+\oV_s t$ in which the equation for the superfluid velocity 
fluctuations becomes
\begin{equation}
\frac{\partial v_s}{\partial t}=-\frac{v_s}{\tau_f}\,,
\label{eq:vs-tau}
\end{equation}
where the friction timescale is
\begin{equation}
\tau_f=\frac{1}{\alpha\kappa L}.
\label{eq:tauf}
\end{equation}

In deriving eq.~(\ref{eq:vs-tau}) we have considered a region of the fluid where the vortex line density, $L$ is essentially constant apart from small fluctuations which cause fluctuations of the friction which are much quicker than the evolution of $v_s$. In this approximation  we can assume that $L$ is constant, so the equation for $v_s$ is linear, and, Fourier trasforming, we have
\begin{equation}
\frac{du_r}{dt}=-\frac{u_r}{\tau_f}\,,
\label{eq:dukdt}
\end{equation}
where $u_r(t)$ is the Fourier component of the superfluid velocity $v_s(x,t)$ at time $t$ and wavenumber $k=2 \pi/r$. The important assumption is that $\tau_f$ does not depend on $k$, so that the linear equation for the velocity $v_s$ in physical space becomes a linear equation for the Fourier component of the velocity $u_r$.

There are two conditions for the existence of the cascade. The first, given above by inequality~(\ref{eq:rggl}), is that the scale $r$ must be larger than the intervortex spacing, that is to say we are in the hydrodynamic regime of many vortices, not in the regime of isolated vortex lines. The second, as for the classical cascade, see condition~(\ref{eq:tau}),  is that
\begin{equation}
\tau_f \gg\tau_r\,.
\label{eq:f}
\end{equation}
Using eq.~(\ref{eq:tauf}) and eq.~(\ref{eq:ur}), which is applicable in the considered hydrodynamic regime of many vortices, eq.~(\ref{eq:f}) becomes
\begin{equation}
\frac{ \epsilon^{1/2} }{ (\alpha \kappa L)^{3/2} } \gg  r\,.
\end{equation}

The energy dissipation rate is
\begin{equation}
\epsilon=-\frac{d}{dt}
\left( 
\frac{1}{V} \int_V \frac{ (\oV_s+v_s)^2}{2}dV \right)
=-\frac{d}{dt} \left( \frac{1}{V} \int_V \frac{v_s^2}{2} dV \right),
\label{eq:dissipation}
\end{equation}
where we have taken into account that the mean value of velocity fluctuations is zero. Using now eqs.~(\ref{eq:vs-tau}) and (\ref{eq:tauf}), the energy dissipation rate can be estimated from eq.~(\ref{eq:dissipation}) as
\begin{equation}
\epsilon=\alpha\kappa L\langle v_s^2\rangle\,,
\label{eq:eps-rms}
\end{equation}
where $\langle v_s^2\rangle$ is the rms of the superfluid velocity fluctuations. We also make the assumption that the average amplitude of the fluctuations, $\ov_s=\langle v_s^2\rangle^{1/2}$ is only a fraction of the mean superflow, $\ov_s = c_1 \vert\oV_s \vert$ with $c_1 \approx 10^{-1}$ or less.

In the counterflow produced by a strong heat current such that the normal fluid is highly turbulent, the turbulent intencity in the latter was recently measured~\cite{Marakov2015} by means of the technique using triplet-state He$^*_2$ molecular tracers. The fluctuations in the normal fluid were found to be about 0.25 of the normal fluid's mean velocity. The fluctuations in the superfluid have not been measured, but the same level of turbulence intencity can be anticipated. However, in the regime where the normal fluid is laminar or stationary, as in the case considered in the current work, velocity fluctuations in the superfluid component should be much smaller.

Since $\oV_s=\rho_n\oV_{ns}/\rho$, combining eqs.~(\ref{eq:gamma}) and (\ref{eq:eps-rms}) we write criteria (\ref{eq:rggl})
and (\ref{eq:f}) in the form
\begin{equation}
c_2 \ell \gg r \gg \ell\,,
\label{eq:conditions}
\end{equation}
where
\begin{equation}
c_2= \frac{\rho_n}{\rho}\frac{c_1}{\alpha \kappa \gamma}\,.
\end{equation}
For temperatures between 1.5 and $2.1\,{\rm K}$ typical of the 
counterflow, $\rho_n/\rho$ is between 0.1 and 0.75 and $\alpha$ 
varies from 0.074 to 0.5 \cite{RJD-CFB-data}, while $\gamma$ is 
of the order of $10^2\,{\rm s/cm^2}$ \cite{Adachi2010}, so that $c_2$ 
is of the order unity and it is impossible to satisfy simultaneously 
both conditions in inequalities (\ref{eq:conditions}). It follows, then, 
that for the values of parameters typical of the thermal counterflow 
the Kolmogorov scaling of the energy spectrum should not be expected. 
(Note that this conclusion has been obtained assuming that the normal 
fluid velocity is constant. This assumption is violated, so that our model 
is no longer valid, if the normal fluid itself becomes turbulent, as in 
the counterflow at a sufficiently large heat current. In this case, owing 
to the mutual friction between the normal fluid and quantized vortices, 
the superfluid energy spectrum acquires the $k^{-5/3}$ Kolmogorov scaling.)

\begin{figure}[ht]
\centering
\includegraphics[width=0.85\linewidth]{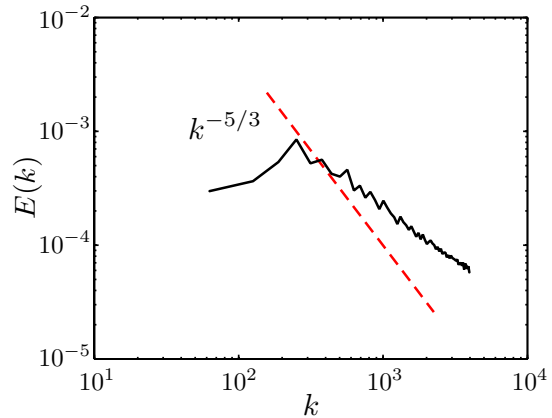}
\caption{{\bf Non-cascading spectrum at high temperature.}
Typical superfluid energy spectrum $E_k$ (arbitrary units) vs 
wave number $k$ (${\rm cm}^{-1}$) in thermal counterflow. 
The hydrodynamical range is from $k_D =63~{\rm cm^{-1}}$ to 
$k_{\ell}=889 ~{\rm cm^{-1}}$. Notice the lack of energy at the 
largest scales and the slope, which is rather different from 
Kolmogorov's $k^{-5/3}$ scaling (indicated by the dashed line).}
\label{fig:counterflow}
\end{figure}

A typical superfluid kinetic energy spectrum obtained from our numerical simulation of the counterflow turbulence is shown in Figure~\ref{fig:counterflow}.

The numerical method and procedure are described in our earlier 
work~\cite{Sherwin2012,Sherwin2015}. The parameters are: temperature $T=1.9~\rm K$, counterflow velocity $V_{ns}= V_n - V_s= 1~\rm cm/s$, friction coefficient $\alpha=0.2$, vortex line density $L \approx 2 \times 10^4~\rm cm^{-2}$ corresponding to $\gamma=141~\rm s/cm^2$, in good agreement with Ref.~\cite{Adachi2010}. The calculation is performed in a cubic periodic domain of size $D=0.1~\rm cm$; numerical discretization along the vortex lines is $\Delta \xi \approx 0.0016~\rm cm$.

As seen from Figure~\ref{fig:counterflow}, the energy spectrum  has a broad peak at intermediate wave numbers, without much energy at the largest scales (smallest $k$). At large $k$, the spectrum follows the typical $k^{-1}$ scaling of smooth isolated vortex lines, not Kolmogorov's $k^{-5/3}$ scaling. In fact, using the parameters of the simulation, inequality~(\ref{eq:conditions}) becomes
$1.8 \ell \gg r \gg \ell$, which cannot be satisfied.
Since the spectrum of an isolated vortex line scales as $k^{-1}$,
the observation of $E_k \sim k^{-1}$ in a turbulent tangle of vortex lines
suggests that far-field effects cancel each other out, in other words that
the vortex lines are randomly oriented, and that the only length scale of the
turbulence is $\ell$.

A similar argument based on the comparison between the time scale of friction with the turnover time of eddies was used to justify the absence of the Kolmogorov energy spectrum in $^3$He-B grid turbulence in the presence of the stationary normal fluid~\cite{Vinen2005}.
On the other hand, the presence of the $k^{-5/3}$ spectrum in $^4$He coflows was clearly demonstrated by the experimental measurements~\cite{Salort2010,Duri2015} and various numerical simulations~\cite{Salort2012,Barenghi-2014-spectra}.

\subsection*{Non-cascading turbulence at low temperatures}

There is strong experimental evidence in both $^4$He and $^3$He-B for the existence of Kolmogorov energy spectra at temperatures so low that the normal fluid, hence the friction, is negligible. This evidence is the observed decay law of the vortex line density, $L \sim t^{-3/2}$, which can be related to an underlying Kolmogorov spectrum. Numerical simulations performed using the Gross-Pitaevskii equation and the Vortex Filament Model confirmed this result by directly measuring the energy spectra.

However, there is also experimental and numerical evidence for a different turbulent regime in which the Kolmogorov spectrum is absent and the resulting decay of turbulence is $L \sim t^{-1}$. As for the high temperature regime discussed in the previous section, the existence of this second regime 
(called {\it Vinen}, or {\it ultraquantum} turbulence in the literature to distinguish it from the {\it quasiclassical} Kolmogorov regime), presents us with a puzzle.

Our aim is to develop a model which would yield conditions necessary for emergence of the quasiclassical quantum turbulence at very low temperatures such that the presemce of the normal fluid can be neglected. A model developed below is based on the experiments reported in Refs.~\cite{Walmsley-2008,Zmeev2015} in which turbulence was generated by injecting a jet of negative ions 
(electrons in a bubble state).
At low temperatures each injected electron dresses itself into a quantized 
vortex ring. The rings then collide and reconnect to produce a vortex tangle which gradually fills the experimental cell. A regime of quantum turbulence was then identified by the scaling with time of the decay of vortex line density after the ion injection has stopped. Numerical analysis of the decay of quasiclassical and ultraquantum regimes of turbulence can be found in our paper~\cite{Baggaley-2012-ultra}. In Refs.~\cite{Walmsley-2008,Zmeev2015} it was found that the regime of turbulence, generated by injection of the ion jet, depends on a duration, $t_i$ of the ion injection: the ultraquantum regime was generated when the duration of injection was relatively short, and the quasiclassical regime was observed if the injection time was longer.

Our model is outlined in the following three paragraphs. 

Assume that vortex rings (whose sizes are narrowly distributed around some mean value) are injected into the experimental cell filled with $^4$He at very low temperature. The rings collide and reconnect, gradually forming the vortex tangle. Part of the energy injected by vortex rings is fed into the Kelvin waves and ultimately dissipated by the phonon radiation. The vortex line density, $L$ grows until, at time $t_*$, called below the ``saturation'' time, the total energy input by injected rings balances the total energy fed into the Kelvin waves. At this time the growth of the line density stops and thereafter the tangle is in the statistically steady state with the time-averaged vortex line density $L_*=L(t_*)$ until time $t=t_i$ when the injection stops and the tangle decays. 

The quasiclassical regime of quantum turbulence may be generated in the case where there exists a mechanism of the three-dimensional inverse energy transfer from the scales at which the energy is injected (presumably, the scale of a single vortex ring) to larger scales which substantially exceed the mean intervortex distance in the created vortex tangle. As we argued in our earlier papers~\cite{Baggaley-2012-ultra,Baggaley-2014-inverse}, such a mechanism can be provided by anisotropy which favours reconnections of loops of the same polarity. Furthemore, in the cited papers we demonstrated by direct numerical simulation a generation of the Kolmogorov $k^{-5/3}$ energy spectrum resulting from the process of rings' injection, thus mimicking the experiments~\cite{Walmsley-2008,Zmeev2015}. The mechanisms of reconnections of injected quantized vortex rings, and of generation of quasiclassical, large-scale velocity fluctuations was further investigated, theoretically and experimentally, in Ref.~\cite{Walmsley-2014}; the authors of cited work favoured the view that the process of rings' reconnections, resulting in creation of the coarse-grained velocity field on the quasiclassical scales, is, in fact, the three-dimensional inverse cascade. However, whether this inverse energy transfer is an inverse cascade or not is irrelevant for the purpose of the current study.

In the process of rings' injection the Kolmogorov spectrum gradually grows through the wavenumber space to smaller wavenumbers eventually filling, at some time $t_K$, the whole available interval of wavenumbers from $k_\ell=2\pi/\ell$ to $k_D=2\pi/D$, where $D$ is the size of the experimental cell (or the integral scale of the fully developed quasiclassical turbulence). If the injection of rings stops at time $t_i\geq t_K$, the following decay of the line density will be that typical of the quasiclassical turbulence and scale with time as $t^{-3/2}$. Had the injection been stopped some time before $t=t_K$, the inverse energy transfer would not have enough time to develop a full Kolmogorov spectrum spanning all available lengthscales, in which case the vortex tangle would still be ultraquantum and decay, after the injection has stopped, as $L\sim t^{-1}$.

Note that our aim is to find conditions necessary for emergence of the quasiclassical regime which is associated with motion at scales larger than the intervortex distance. For such a motion the rate at which the energy is fed into the Kelvin waves plays a r\^ole of the dissipation rate, and the details of dissipation by phonon emission as well as the possibility of the much debated bottleneck in the energy flux between the macroscopic motion and the Kelvin wave cascade are irrelevant.

The energy of an isolated quantized vortex ring of radius $R$ 
is~\cite{Donnelly} 
\begin{equation}
E_r=\frac{\rho\kappa^2R}{2}F\,,
\label{eq:Er}
\end{equation}
where the function $F=F(R)$ is
\begin{equation}
F=\ln{\left( \frac{8R}{a_0} \right)} -\frac{3}{2}\,,
\label{eq:F}
\end{equation}
$a_0\approx0.1\,{\rm nm}$ is the vortex core radius which is of the order of coherence length. We assume here that all injected rings are of the same radius, as practically was the case in the experiments~\cite{Walmsley-2008,Zmeev2015}. (In fact, the sizes of the rings are narrowly distributed around some value of the radius. This was taken into account in our earlier numerical studies~\cite{Baggaley-2012-ultra,Baggaley-2014-inverse}, but is irrelevant for the purpose of the model considered below.)

Assuming that the frequency of rings' injection is $f_{\rm in}$, the rate of energy input, per unit mass, can be estimated as
\begin{equation}
e_{\rm in}=\frac{1}{2V}\kappa^2 f_{\rm in}RF(1-p)\,,
\label{eq:ein}
\end{equation}
where $V=D^3$ is the volume of cubic experimental cell, and $p=p(L(t),\,D)$ is the probability of the vortex ring to propagate through the vortex tangle without collisions with other vortex lines. This probability, which depends on the current vortex line density, $L(t)$ and the size of experimental cell, $D$ was investigated in detail in Ref.~\cite{Laurie-2015}. The probability $p$ is close to unity when the tangle is still very dilute but decreases exponentially with both the vortex line density and the size of experimental cell. For the values of parameters typical of the experiment~\cite{Walmsley-2008} ($R\approx5.3\times10^{-5}\,{\rm cm}$ and $D=4.5\,{\rm cm}$), the probability $p$ becomes small (less than 0.25) already for $L\approx3\times10^3\,{\rm cm}^{-2}$, and negligible when the tangle reaches the saturated, statistically steady state with $L$ of the order $10^4\,{\rm cm}^{-2}$.

For the sake of simplicity we assume below that $p=0$ so that all injected rings contribute to the growth of the line density. (Note that in the conditions of the cited experiment the probability $p$ is not unity even at the very early stages of tangle's formation as the rings are injected in a narrow beam so that collisions between rings of slightly different radii are not infrequent. This can also be seen in our earlier numerical simulations~\cite{Baggaley-2012-ultra,Baggaley-2014-inverse} of the experiment~\cite{Walmsley-2008}.) The saturation time, $t_*$ required for the tangle to reach the statistically steady state will, therefore, be somewhat underestimated. However, as will be seen below, for the parameters of the experiment~\cite{Walmsley-2008} a formation of the large scale motion (and, therefore, of the quasiclassical tangle) occurs mainly after the statistically steady state has been reached, that is at times $t>t_*$. Moreover, it will also be shown below that time $t_K$ required for generation of quasiclassical turbulence is independent of $t_*$.
We will also assume that at each moment of time the vortex tangle occupies the whole experimental volume. Although there is some evidence that at the early stages of evolution the tangle occupies only part of the experimental volume, see e.g. Fig.~1 in Ref.~\cite{Walmsley-2008}, 
the simplifying assumptions made above will suffice for the order-of-magnitude estimates of $t_K$.

The rate, per unit mass, at which energy is fed into the Kelvin waves, and ultimately dissipated, is given by~\cite{VinenNiemela}
\begin{equation}
\epsilon=G\kappa^3\ell^{-4}=G\kappa^3L_0^2\,,
\label{eq:KWflux}
\end{equation}
where $L_0$ is the \textit{smoothed} vortex line density~\cite{VinenNiemela}, that is the length of line per unit volume after the excited Kelvin waves 
have been removed, and $G$ is a constant whose numerical value will be 
discussed below.

Before proceeding with our model, we have to make a rather important remark. 
In the zero-temperature quantum turbulence the vortex reconnections 
generate the Kelvin waves which lead to an increase of the \textit{actual} 
(as opposed to the \textit{smoothed}) vortex line density, $L$, for which 
the relation~(\ref{eq:KWflux}) still holds in the form 
$\epsilon=G'\kappa^3L^2$, but with $G'<G$. 
The product $G\kappa$ can be interpreted as the effective viscosity, 
$\nu'$ which was thoroughly discussed in 
Refs.~\cite{VinenNiemela,Walmsley-2008,Walmsley-2007}. 
Note also that the mean intervortex distance, $\ell$ is linked by the relation $\ell\approx L_0^{-1/2}$ to the smoothed line density, not to the actual vortex line density $L$.

In the last two decades significant experimental 
\cite{Stalp1999,Walmsley-2008,Walmsley-2007} and theoretical/numerical 
\cite{Tsubota2000,Kondaurova2014} efforts were made to determine the 
value of the effective kinematic viscosity in the zero-temperature 
turbulent $^4$He. In particular, it was argued~\cite{Walmsley-2007} 
that the value of $\nu'$ (or $G$) should depend on the spectrum of quantum 
turbulence and, therefore, its value for the quasiclassical (Kolmogorov) 
regime should differ from that for the ultraquantum (Vinen) turbulence. 
However, the recent work~\cite{Zmeev2015} gives convincing arguments 
that the value of $G$ is independent of the energy spectrum and, 
therefore, should be the same for both the quasiclassical and the 
ultraquantum regimes. 
Earlier experiments~\cite{Stalp1999,Walmsley-2008,Walmsley-2007} and 
theoretical/numerical results~\cite{Tsubota2000,Kondaurova2014} 
suggested the value of $G$ in the interval 0.06--0.10 (although in some 
cases it is not clear whether $G$ or $G'$ was actually measured). 
A very recent study~\cite{Zmeev2015} suggests that, 
in the zero-temperature limit, $G \approx 0.08$ (a more precise 
estimate for $G$ is hardly possible at present). As will be seen below, 
for our order-of-magnitude estimates of time $t_K$ 
required for the formation of the quasiclassical quantum 
turbulence a precise value of $G$, and even the uncertainty whether the 
value used for our estimates is that of $G$ or of $G'$ are unimportant, 
and, following Ref.~\cite{Zmeev2015}, we will assume that $G=0.08$.

Using the assumption that $p=0$, we model the growth of the smoothed vortex line density during the tangle's formation by the equation
\begin{equation}
\frac{dL_0}{dt}=\beta\frac{2\pi Rf_{\rm in}}{V}\,.
\label{eq:Lgrowthrate}
\end{equation}
Here we introduced the empirical dimensionless constant $\beta<1$ to account for several phenomena. Firstly, reconnection of the vortex loop, whose line length is $L_R=2\pi R$, with another vortex ring or with a vortex line within the tangle does not, in general, increase the tangle's total {\it smoothed} line length by $L_R$ as some of the length may be ``lost'' to the Kelvin waves (for example, a merger of two rings of the same radius $R$ results in the ring whose smoothed lengths corresponds to a radius of only about $\sqrt{2}R$). In fact, $\beta$ should be a function of the vortex line density, and, secondly, should also incorporate a dependence on the probability $p$ for the ring to propagate through the tangle without collisions. However, a somewhat simplistic model given by eq.~(\ref{eq:Lgrowthrate}) with $\beta={\rm constant}$ will suffice: as will be seen below, the formation of the full Komogorov spectrum occurs after saturation of the tangle (at least for the parameters corresponding to experiment~\cite{Walmsley-2008}). Moreover, neither the saturated line density, $L_*$, nor the time $t_K$ of the formation of quasiclassical turbulence depend on the parameter $\beta$.
For our estimates of time $t_K$ we will assume that $0.5<\beta<1$.

Neglecting a small vortex line density at the beginning of the rings' injection so that $L_0(0)=0$, we integrate eq.~(\ref{eq:Lgrowthrate}) within the period of tangle saturation, so that
\begin{equation}
L_0=2\pi\beta Rf_{\rm in}t/V\,,
\label{eq:Lt}
\end{equation}
until $t=t_*$ defined such that
\begin{equation}
E_{\rm in}(t_*)=E_{\rm diss}(t_*)\,,
\label{eq:condition-tstar}
\end{equation}
where
\begin{equation}
E_{\rm in}(t)=\int_0^t e_{\rm in}\,dt=\frac{\kappa^2f_{\rm in}RF}{2V}\,t
\label{eq:Ertot}
\end{equation}
and
\begin{equation}
E_{\rm diss}=\int_0^t \epsilon(t)\,dt=G\kappa^3\int_0^tL_0^2(t)\,dt=  \frac{(2\pi)^2}{3}\,\frac{\beta^2G\kappa^3R^2f_{\rm in}^2}{V^2}\,t^3
\label{eq:Edtot}
\end{equation}
are, respectively, the total injected energy and the total energy dissipated by the Kelvin wave cascade (both quantities are per unit mass). The total dissipation, $E_{\rm diss}$, which grows with time as $t^3$, cannod exceed the total energy input $E_{\rm in}$ whose growth with time is linear. We, therefore, assume that at time $t=t_*$ following from eqs.~(\ref{eq:condition-tstar})-(\ref{eq:Edtot}) in the form
\begin{equation} 
t_*^2=\frac{3VF}{8\pi^2\beta^2G\kappa f_{\rm in}R}\,,
\label{eq:t*}
\end{equation}
the growth of $L_0$ stops and hereafter
\begin{equation}
L_0=L_0(t_*)=L_*, \quad \ell=\ell_*=L_*^{-1/2}, \quad {\rm and} \quad \epsilon=\epsilon_*=G\kappa L_*^2\,,
\label{eq:Lstar}
\end{equation}
where
\begin{equation}
L_*=\left(\frac{3f_{\rm in}RF}{2G\kappa V}\right)^{1/2}\,, \quad \ell_*=\left(\frac{2G\kappa V}{3f_{\rm in}RF}\right)^{1/4}\,, \quad \epsilon_*=\frac{3\kappa^2f_{\rm in}RF}{2V}
\label{eq:stars}
\end{equation}
(note that these quantities do not depend on the modelling parameter $\beta$ in eq.~(\ref{eq:Lgrowthrate}).

We turn now to the formation of quasiclassical turbulence. We assume that, resulting from the inverse energy transfer (cascade) induced by reconnections between injected vortex loops, all injected energy gradually forms a quasiclassical tangle up to wavenumber $k$ which depends on time. We also assume that during this process the quasiclassical turbulence remains quasistationary. The energy, per unit mass, of the quasiclassical tangle is then given by
\begin{equation}
E_K=\int_{k(t)}^{k_\ell}C_K\epsilon^{2/3}k^{-5/3}\,dk=\frac{3}{2}C_K\epsilon^{2/3}\left(k^{-2/3}-k_\ell^{-2/3}\right)\,,
\label{eq:Ekt}
\end{equation}
where $k_\ell=2\pi/\ell(t)=2\pi L_0^{1/2}(t)$ and $C_K\approx1.5$ is the Kolmogorov constant.

Quasiclassical turbulence is dissipated at the scale of the intervortex distance, $\ell(t)$ by the Kelvin wave cascade, so that the dissipation in eq.~(\ref{eq:Ekt}) is given by eq.~(\ref{eq:KWflux}) with $L_0(t)$ determined by eq.~(\ref{eq:Lt}). From the equation
\begin{equation}
E_{\rm in}(t)=E_K(t)\,,
\label{eq:EinEK}
\end{equation}
where $E_{\rm in}$ grows in time as in eq.~(\ref{eq:Ertot}), we can extract $k(t)$, and the corresponding ``quasiclassical'' lengthscale $\lambda=2\pi/k$:
\begin{equation}
\lambda(t)=\ell(t)\left[1+\frac{2(2\pi)^{2/3}}{3C_K}\,\frac{E_{\rm in}(t)}{\epsilon^{2/3}(t)\ell^{2/3}(t)}\right]^{3/2}\,,
\label{eq:lambda}
\end{equation}
where $\ell(t)=L_0^{-1/2}$ 
and $\epsilon(t)$ is determined by eq.~(\ref{eq:KWflux}) (for $t>t_*$ these quantities no longer depend on time and are determined by formulae~(\ref{eq:stars})). The lengthscale $\lambda$ should be compared to $\ell$ and the size $D$ of the experimental cell. We expect that if $\lambda/\ell\approx10$ (one decade of the $k$-space) the Kolmogorov spectrum should start becoming visible. The full Kolmogorov spectrum, which would yield the observed $L\sim t^{-3/2}$ decay of the vortex line density after the injection of rings has stopped, requires, of course, $\lambda(t)$ to grow up to the largest possible value, $\lambda(t_K)=D$.

We should note here that eq.~(\ref{eq:lambda}) is not applicable during a very short time period after the start of injection when the vortex configuration still consists mainly of small individual vortex loops rather than of tangled vortex lines, so that eq.~(\ref{eq:KWflux}) is not yet valid. However, this short time period will not significantly affect our estimates of times $t_*$ and $t_K$.

We now analyze time $t_K$ predicted by eq.~(\ref{eq:lambda}) in connection with experimental observations. Two experiments in which the quasiclassical regime of quantum turbulence was generated by ion injection were reported in Refs.~\cite{Walmsley-2008} and \cite{Zmeev2015}. In the first of these experiments, performed at temperatures ranging from $T=0.7\,{\rm K}$ to $T=1.6\,{\rm K}$ in a cube-shaped container with sides 4.5 cm (volume $V\approx91\,{\rm cm}^3$), the quasiclassical regime was prominent after injections longer than 30~s, depending on the injected electron current $I$ which was in the range from $10^{-12}$ to $10^{-10}\,{\rm A}$. In the second experiment~\cite{Zmeev2015}, which was performed in the experimental cell of a more complicated shape, the quasiclassical regime has been observed after longer, $\sim100\,{\rm s}$ injection (a detailed description of the experimental cell can be found in Ref.~\cite{Zmeev2014}). Note that Refs.~\cite{Walmsley-2008,Zmeev2015} do not provide a more detailed analysis of the relation between the duration of injection and the resulting regime of turbulence. Although the second of these experiments has been carried out in the truly zero-temperature limit (at $T=80\,{\rm mK}$), to illustrate our calculation of time $t_K$ we will use the parameters of the first experiment~\cite{Walmsley-2008} (assuming, however, $T=0$) whose geometry of the experimental cell was much simpler.

\begin{figure}[ht]
\centering
\includegraphics[width=0.85\linewidth]{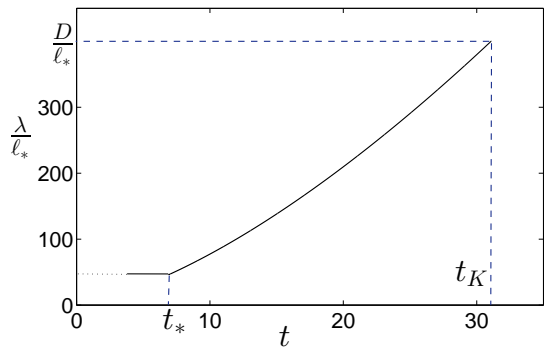}
\caption{{\bf Quasiclassical lengthscale $\lambda$ (in units of $\ell_*$) 
vs time (in s).} Here $t_K$ is the time when $\lambda$ has become equal 
to the experimental cell's size so that the full Kolmogorov spectrum is 
formed if the injection time $t_i\geq t_K$; $t_*$ is the saturation time. 
Dotted part of the curve corresponds to the short time period after the 
beginning of injection when the vortex configuration still consists of 
individual loops and $\ell$ cannot yet be defined.
See text for details and values of parameters used in calculation.}
\label{fig:lambda}
\end{figure}

In experiment~\cite{Walmsley-2008}, the radius of injected vortex rings was $R\approx0.53\,\mu{\rm m}=5.3\times10^{-5}\,{\rm cm}$, which corresponds to $F\approx9.5$. To find the frequency of rings' injection, we will 
follow the assumption made in Ref.~\cite{Walmsley-2008} that at low 
temperatures each injected electron dresses itself in a quantized 
vortex ring. Then, $f_{\rm in}=I/e$, where $e\approx1.6\times10^{-19}\,{\rm C}$ is the elementary charge. For the electron current $I=10^{-10}\,{\rm A}$ the frequency $f_{\rm in}=6.25\times10^8\,{\rm Hz}$, and, assuming $\beta=0.5$, for the saturation time we have $t_*\approx6.92\,{\rm s}$.
For the parameters of experiment~\cite{Walmsley-2008}, the evolution of 
the quasiclassical lengthscale, $\lambda(t)$ calculated using 
formula~(\ref{eq:lambda}) is shown in Figure~\ref{fig:lambda}. 

Shortly after the beginning of injection the ratio $\lambda(t)/\ell(t)$ reaches the value
\begin{equation}
\frac{\lambda(t)}{\ell(t)}=\left[1+\frac{2(2\pi)^{2/3}}{9C_K}\,\frac{\epsilon_*^{1/3}t_*}{\ell_*^{2/3}}\right]^{3/2}\,,
\label{eq:lambda-ell}
\end{equation}
which remains constant during the period $0<t<t_*$ while the line density still grows and $\ell$ decreases. (Note that solution~(\ref{eq:lambda}) is not valid for a short period of time immediately after beginning of injection when the vortex configuration still consists of individual vortex loops and $\ell$ cannot yet be defined; for this time period solution~(\ref{eq:lambda}) is shown in Figure~\ref{fig:lambda} by the dotted part of the curve.) The formation of quasiclassical scales occurs mainly after the tangle has been saturated, that is for $t>t_*$ (at which point $\lambda/\ell=46$) when, as shown in Figure~\ref{fig:lambda}, the quasiclassical lengthscale increases rapidly as $\lambda(t)\sim t^{3/2}$:
\begin{equation}
\lambda(t)=\ell_*\left[1+\frac{2(2\pi)^{2/3}}{9C_K}\,\frac{\epsilon_*^{1/3}t}{\ell_*^{2/3}}\right]^{3/2}\,.
\label{eq:lambda-increase}
\end{equation}
In eqs.~(\ref{eq:lambda-ell})-(\ref{eq:lambda-increase}), $t_*$, $\ell_*$ and $\epsilon_*$ are determined by formulae~(\ref{eq:t*}) and (\ref{eq:stars}).

Since, for the parameters of experiment~\cite{Walmsley-2008}, $t_K>t_*$, then the time when $\lambda$ becomes equal to the cell's size so that the quasiclassical turbulence is fully developed can be calculated from eq.~(\ref{eq:lambda}), assuming $\lambda=D$, $\ell=\ell_*$, and $\epsilon=\epsilon_*$, as
\begin{equation}
t_K=\frac{9C_K}{2(2\pi)^{2/3}}\,\frac{\ell_*^{2/3}}{\epsilon_*^{1/3}}\left[\left(\frac{D}{\ell_*}\right)^{2/3}-1\right]\,.
\label{eq:tK}
\end{equation}
Note that time $t_K$ does not depend on the modelling parameter $\beta$ in eq.~(\ref{eq:Lgrowthrate}). For the parameters of experiment~\cite{Walmsley-2008} eq.~(\ref{eq:tK}) yields $t_K\approx31\,{\rm s}$.

Although this time compares very favourably with the results of experiment~\cite{Walmsley-2008} which showed that at temperatures $T\geq0.7\,{\rm K}$ the quasiclassical regime is especially prominent after more than 30~s long injection, such a good agreement might be rather coincidental. The reason is that even at temperature as low as 0.7~K, such that the normal fluid fraction is only $2.27\times10^{-4}$ and the mutual friction coefficient $\alpha$ is of the order of $10^{-3}$, the vortex rings of radius $0.53\,\mu{\rm m}$ cannot be treated as ballistic. Indeed, at this temperature such rings decay on a distance $R/\alpha\approx0.05\,{\rm cm}$ which is much smaller than the size of the experimental cell. Only at temperatures $T<0.5\,{\rm K}$ ($\alpha<10^{-5}$) the range of rings' decay exceeds $D$. It can be expected that, in the experiment~\cite{Walmsley-2008}, at the initial stage of ion injection the formation of tangle and the generation of motion on quasiclassical ($r>\ell$) scales occur near the injector. It can also be expected that, as time progresses, the tangle and the flow on quasiclassical scales spread through the whole experimental cell, as illustrated by the cartoon shown in Fig.~1 of Ref.~\cite{Walmsley-2008}. Clearly, our simple model, which assumes both ballistic propagation of rings and the spatial uniformity of the tangle, does not capture these phenomena. However, because our model captures essential features of the generation of motion on  quasiclassical scales, we should probably expect an order-of-magnitude agreement between our prediction of time $t_K$ and the experimental observations. A better agreement found in this work is somewhat surprising.

\begin{figure}[ht]
\centering
\includegraphics[width=0.85\linewidth]{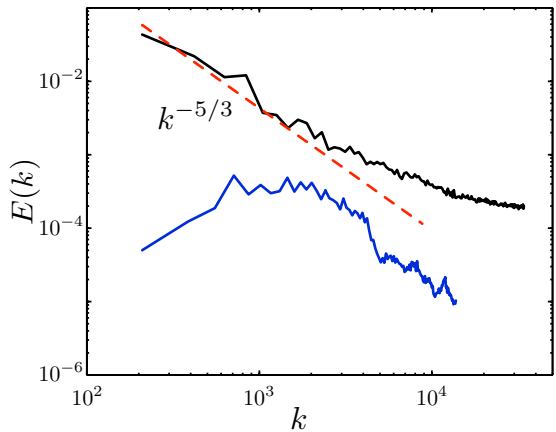}
\caption{{\bf Cascading and non-cascading spectrum at low temperature.}
Energy spectrum (arbitrary units) vs wave number (${\rm cm}^{-1}$) for 
the tangle numerically generated by injection of the vortex rings. 
Bottom: duration of injection $t_i=0.1\,{\rm s}$. 
Top: prolonged injection for $t_i=1.0\,{\rm s}$. 
See text and Ref.~\cite{Baggaley-2012-ultra} for details and values of parameters used in calculation.}
\label{fig:quasi-ultra}
\end{figure}

Obtained from our numerical simulation~\cite{Baggaley-2012-ultra}, typical energy spectra for ultraquantum and quasiclassical regimes of quantum turbulence, generated, respectively, by a short ($t_i=0.1$~s) and a long ($t_i=1$~s) injection of vortex ring, is illustrated in
Figure~\ref{fig:quasi-ultra}. The calculation was performed in the periodic box of size $D=0.03\,{\rm cm}$ for vortex rings, injected with initial velocities randomly confined within a small, $\pi/10$ angle, of radii narrowly distributed arond $R=6\times10^{-4}\,{\rm cm}$. The frequency of rings' injection was $f_{\rm in}\approx 320\,{\rm Hz}$ (see Ref.~\cite{Baggaley-2012-ultra} for
numerical method and procedure).
As seen from the bottom curve, in the case of short duration 
of the injection the spectrum does indeed show the 
absence of the $k^{-5/3}$ scaling for any interval of wave numbers 
but has instead a broad peak similar to that typical of the counterflow 
turbulence, cf. Figure~\ref{fig:counterflow}. On the other hand, 
our simulation for a longer duration of injection 
clearly shows that in this case the Kolmogorov spectrum is formed, 
see the top curve of Figure~\ref{fig:quasi-ultra}.

\subsection*{Discussion}

Based on estimates for the effective dissipation rate in quantum turbulence 
considered at the hydrodynamic scales, that is in the range of scales 
corresponding to many vortices rather than individual vortex lines, we 
have obtained conditions necessary for existence of the kinetic energy cascade 
in the superfluid component and, hence, of the $k^{-5/3}$ scaling of the 
superfluid's kinetic energy spectrum. The phenomenological approach which 
we have developed has enabled us to explain why, at finite temperatures, 
the Kolmogorov energy spectrum cannot be observed in $^4$He counterflows 
and $^3$He-B flows in the presence of stationary normal fluid. We have then 
extended our approach to consider generation of $^4$He quantum turbulence 
at very low temperatures. We have considered a tangle of quantized vortices 
generated, as in the experiments~\cite{Walmsley-2008,Zmeev2015}, by a beam of electrons 
injected, in the bubble state, into helium at temperatures significantly 
lower than $T_\lambda$. At the considered scales of many vortices the r\^ole 
of the effective dissipation rate is played by the rate at which the 
energy is fed into the Kelvin waves. Having calculated the time required for the inverse energy transfer to form the Kolmogorov energy spectrum for all available ``quasiclassical'' wavenumbers (from that corresponding to the size of experimental volume to the wavenumber corresponding to the intervortex distance $\ell$), we
estimated the durations of injection required to generate the quasiclassical regime of quantum turbulence at scales larger than $\ell$. The calculated durations are consistent with those 
observed experimentally~\cite{Walmsley-2008}.

We conclude that the regimes of quantum turbulence which have been observed 
at both high and low temperatures and which are characterized by the spectral nature in which the larges eddies are weak, most of the energy is contained at the intermediate scales, and the fully developed energy cascade is absent,  
can be understood on the ground of simple large-scale quasiclassical
(``hydrodynamical'') considerations.

Data supporting this publication is openly available under an 
%Open Data Commons Open Database License \cite{data}.
Open Data Commons Open Database License.

\section*{Acknowledgements}

We acknowledge the support by the Engineering and Physical Sciences 
Research Council (United Kingdom) through Grant No.~EP/I019413/1.

\end{document}